\documentclass[5p]{elsarticle}
\journal{Physics Letters B} 

\newlength{\figw} 
\usepackage{notoccite}

\usepackage{graphicx,epsf, epsfig} 
\usepackage{bm}
\usepackage{longtable,tabularx} 
\usepackage{color}
\usepackage[breaklinks]{hyperref}
\usepackage{amsfonts,amsmath,amssymb,mathrsfs,gensymb}
\usepackage{array} 
\usepackage{multirow}
\usepackage{rotating,array} 
\usepackage{soul}
\usepackage[normalem]{ulem}

\definecolor{linkcolor}{rgb}{0.0,0.3,0.5}
\definecolor{venetianred}{rgb}{0.78, 0.03, 0.08}
\hypersetup{colorlinks=true,linkcolor=linkcolor,citecolor=linkcolor,filecolor=linkcolor,urlcolor=linkcolor}

\newcommand{\lp}{\ensuremath{\left(}}
\newcommand{\rp}{\ensuremath{\right)}}

\newcommand{\ex}{\mbox{e}} \newcommand{\dd}{\mbox{d}}

\newcommand{\ci}{\mathsf{i}}

\begin{document}

\title{Radial excitations of current-carrying vortices}

\author[B]{Betti Hartmann}
\ead{bhartmann@ifsc.usp.br}
\address[B]{Instituto de F\'isica de S\~ao Carlos (IFSC), Universidade de S\~ao
Paulo (USP), CP 369,  13560-970, S\~ao Carlos, SP, Brasil}

\author[F]{Florent Michel} 
\ead{florent.michel@th.u-psud.fr}
\address[F]{Laboratoire de Physique
Th\'eorique, CNRS, Univ. Paris-Sud, Universit\'e Paris-Saclay, 91405
Orsay, France}

\author[P]{Patrick Peter} 
\ead{peter@iap.fr}
\address[P]{Institut d'Astrophysique de Paris
($\mathcal{G} \mathbb{R} \varepsilon \mathbb{C} \mathcal{O}$), UMR 7095
CNRS, Sorbonne Universit\'es, UPMC Universit\'e Paris 6,\\ Institut Lagrange de
Paris, 98 bis boulevard Arago, 75014 Paris, France.}

%%\pacs{\vspace{-0.15cm}  98.80.Cq, 11.27.+d, 04.40.Nr }

\date{\today}

\begin{abstract}
We report on the existence of a new type of cosmic string solutions  in
the Witten model with $\mathrm{U}(1) \times \mathrm{U}(1)$ symmetry.
These solutions are superconducting with radially excited condensates
that exist for both gauge and ungauged currents. Our results suggest
that these new configurations can be macroscopically stable, but
microscopically unstable to radial perturbations. Nevertheless, they
might have important consequences for the network evolution and
particle emission. We discuss these effects and their possible
signatures. We also comment on analogies with non-relativistic
condensed matter systems where these solutions may be observable.
\end{abstract}

\maketitle

\section*{Introduction}

Linear topological defects, e.g., vortex lines in condensed matter
physics \cite{Vortices,Abrikosov}, superfluid vortices (see e.g.
\cite{superfluid}), non-Abelian vortices (see e.g. \cite{shifman}),
also frequently appear as solutions of high energy field theories where
they are called cosmic strings \cite{VS}; these include grand unified
(GUT) or superstring theories. Such strings yield many cosmological and
astrophysical consequences \cite{Ade:2013xla,Vachaspati:2015cma} that
have not all been yet fully investigated. Many microscopic models have
been discussed, providing direct contact with the underlying field
theory \cite{Allys}.

Even in the simplest models such as the popular $\mathrm{U}(1)$ Abelian
Higgs model with a Mexican hat potential, there are no known analytical
solutions, straight and static ones \cite{no} having only been
constructed numerically. These so-called {\it Abelian--Higgs} or {\it
Abrikosov-Nielsen--Olesen} (ANO) vortices have a localized
energy-momentum tensor and a quantized ma\-gne\-tic-like flux with the
magnetic-like field pointing in the direction of the string axis.
Numerical solutions were found in Refs.~\cite{vega_schapo,gl,lm} and
their gravitational effects have been studied in
Refs.~\cite{gl,lm,gar,linet,cg,clv,bl}.

The ANO string and flux tube widths are inversely proportional,
respectively, to the Higgs and the gauge boson masses. When these
masses are equal, the strings saturate the
Bogomolonyi--Prasad--Sommerfield (BPS) bound \cite{bogo} implying that
the energy per unit length is proportional to the topological charge.
This, in turn, guarantees stability; the fields then satisfy a set of
coupled first order differential equations whose solutions are also not
analytically known and thus have to be constructed numerically.
Furthermore, it was shown that BPS strings remain so in curved
space-time, i.e., they satisfy the same energy bound as in flat
space-time \cite{linet,cg}.

Cosmic strings however can be, and more often than not are
\cite{Davis:1995kk} {\it superconducting} \cite{witten}, carrying
persistent currents that can be spacelike (like an ordinary current),
timelike (akin to a charge), or null (chiral or lightlike
\cite{bcpp99}). Currents up to $10^{20}$ Amps (GUT case) can be induced
by either bosonic or fermionic \cite{CR01} charge carriers. In the
former case, the scalar field that undergoes spontaneous symmetry
breaking, leading to the formation of the strings, couples
non-trivially to a second one. For appropriate choices of the
self-couplings and vacuum expectation values of the scalar fields, the
second field can form a condensate in the core of the string.

While standard cosmic strings have energy per unit length $\varepsilon$
equal to their tension $\mu$, this ceases to be true in the presence of
a current. The relation between the energy per unit length and the
tension of a superconducting string solution of the
$\mathrm{U}(1)\times \mathrm{U}(1)$ model has been discussed in detail
in \cite{Peter:1992dw,Peter:1992sz} and it has been suggested that the
equation of state is of logarithmic form \cite{Carter:1994hn}. This has
been confirmed numerically in \cite{hartmann_carter}. Superconducting
solutions also exist in other models, such as the semilocal
$\mathrm{SU}(2)_{\rm global}\times \mathrm{U}(1)_{\rm local}$ model
\cite{Abraham:1992hv, Forgacs_volkov} as well as in the
$\mathrm{SU}(2)_{\rm local}\times \mathrm{U}(1)_{\rm local}$
electroweak model \cite{Garaud}.

In this Letter we report on a new type of solutions in the
$\mathrm{U}(1)\times \mathrm{U}(1)$ model: superconducting strings with
radially excited condensates. These solutions possess a finite number
of nodes in the scalar field function associated to the unbroken
$\mathrm{U}(1)$ symmetry. Radial excitations of solitonic-like
solutions are quite well known: they appear in non-topological soliton
systems such as $Q$-balls and boson stars \cite{kkl} with an unbroken
symmetry as well as in topological soliton systems such as magnetic
monopoles \cite{bfm} in which a continuous symmetry gets spontaneously
broken. To our knowledge they were so far never considered in the
present context. Yet, studying the solutions of the
$\mathrm{U}(1)\times \mathrm{U}(1)$ model using both analytical and
numerical techniques, we find that they are a generic prediction. They
are thus important to understand the mathematical structure of the
theory, and may have nontrivial consequences for the evolution of the
string network as well as the emission of particles during reconnection
between two strings.

\section{The model}

We study the Witten model of superconducting strings with bosonic
currents \cite{witten}. This model contains two complex scalar fields
$\phi$ and $\sigma$ which are minimally coupled to two (different) U(1)
gauge fields. Using a metric with signature $(+,-,-,-)$, the Lagrangian
density reads
\begin{align}
\mathcal{L} =& -\frac{1}{4} F_{\mu\nu}F^{\mu\nu} -\frac{1}{4}
G_{\mu\nu} G^{\mu\nu} + \frac{1}{2} D_{\mu} \phi \left(D^{\mu}
\phi\right)^{*} \nonumber \\
& + \frac{1}{2} D_{\mu} \sigma \left(D^{\mu}
\sigma\right)^{*}-V(\phi,\sigma),
\label{lag}
\end{align}
where the potential is given by
\begin{equation}
V(\phi,\sigma)= \frac{\lambda_1}{4}(\vert\phi\vert^2-\eta_1^2)^2
+\frac{\lambda_2}{4}\vert\sigma\vert^2(\vert\sigma\vert^2-2\eta_2^2)
+\frac{\lambda_3}{2}\vert\phi\vert^2 \vert\sigma\vert^2  ,
\label{pot}
\end{equation}
the gauge covariant derivatives read
\begin{equation}
D_\mu \phi = \left(\partial_\mu - \ci \, e_\phi B_\mu\right) \phi \ \ , \ \  
D_\mu \sigma = \left(\partial_\mu - \ci \, e_\sigma A_\mu\right) \sigma \ ,
\end{equation}
and the field strength tensors of the two U(1) gauge fields $A_{\mu}$
and $B_{\mu}$ are
\begin{equation}
 G_{\mu\nu}=\partial_{\mu} B_{\nu} - \partial_{\nu} B_{\mu} \ \ , \ \
 F_{\mu\nu}=\partial_{\mu} A_{\nu} - \partial_{\nu} A_{\mu} \ .
\end{equation}
\\
In the following we will use cylindrical coordinates $(t,r,\theta,z)$ and work 
with the ansatz
\begin{align}
& B_{\mu} \dd x^{\mu} = \frac{1}{e_\phi} \left[ n - P\left(r\right)
\right] \dd\theta \  ,  \ \phi(r,\theta)=\eta_1 h(r) \, \ex^{\ci
n\theta} \ , \ \nonumber \\
& A_\mu \dd x^\mu = A_z(r) \dd z + A_t(r) \dd t \ ,\
\sigma(t,r,z)=\eta_1 f(r) \, \ex^{\ci(\omega t-kz)}.
\end{align}
From now on, we shall restrict attention to the case $e_\sigma=0$, for
which the external gauge field $A_\mu$ can be set to zero as well; in
fact, this gauge field is sourced by the current flowing along the
string but hardly backreacts on the string microstructure
\cite{Peter:1992sz}. In Section~\ref{sec:discussion} we briefly comment
on the case $e_\sigma \neq 0$.

Introducing the dimensionless coordinate $x := \sqrt{\lambda_1} \eta_1
r$, the equations of motion only depend on the dimensionless coupling
constants
\begin{equation}
\alpha^2=\frac{e_\phi^2}{\lambda_1} \ , \ q=\frac{\eta_2}{\eta_1} \  ,
\ \gamma_i = \frac{\lambda_i}{\lambda_1} \ \ (i=2,3),
\end{equation}
as well as the (also dimensionless) state parameter
$w:=(k^2-\omega^2)/(\lambda_1\eta_1^2)$; they read
\begin{align}
\left(\frac{P'}{x}\right)'
&= \alpha^2 \frac{P h^2}{x} \ , \label{gauge} \\ \frac{1}{x}\left( x
h'\right)'
&= h\left( \frac{P^2}{x^2} + h^2 - 1 + \gamma_3 f^2 \right) \ ,
\label{higgs} \\ \frac{1}{x}\left( x f'\right)'
&= f\left[ w  + \gamma_2  \left(f^2 - q^2\right) + \gamma_3  
h^2\right], 
\label{condensate}
\end{align}
where a prime denotes a derivative with respect to $x$. The boundary
conditions at the string axis $x=0$ and at infinity read
\begin{align} 
P(x)\vert_{x=0} & = n \ , \ h(x)\vert_{x=0}=0 \ , \ f'(x)\vert_{x=0}=0
\ , \nonumber \\ \lim_{x\rightarrow \infty} P(x) & = 0 \ ,
\lim_{x\rightarrow \infty} h(x) = 1 \ , f(x) \mathop{=}_{x \to \infty}
o \lp x^{-1/2} \rp
\label{BoundCond}
\end{align} 
The important physical quantities of the solutions are the energy per
unit length $\varepsilon$ and the tension $\mu$, given (in rescaled
units) by
\begin{equation}
\varepsilon= -2\pi \int x T^t_t \, \dd x  \ \ , \ \ \mu=-2\pi \int x
T^z_z \, \dd x \ ,
\end{equation}
respectively, with the energy-momentum tensor
\begin{equation}
 T_{\mu\nu}=-2 \frac{\delta \mathcal{L}}{\delta g^{\mu\nu}} +
 g_{\mu\nu} \mathcal{L} \ ,
\end{equation}
as well as the (rescaled) absolute value $I$ of the current~:
\begin{equation}
I=2\pi \sqrt{\vert w\vert} \int x
f^2\, \dd x   \ .
\end{equation}

Note that our model bears strong similarities with that of Ref.
\cite{extended} provided one makes the replacements $\eta_1 \to 1$,
$\beta_1 \to 2 \lambda_1$, $ \beta_2 \to \lambda_2/2$, $\beta' \to
\lambda_3/2$ and $\alpha \to \lambda_2 \eta_2^2/2$; it was shown that
different types of solutions may exist in this model, but for the
purpose of exhibiting our new configuration, we shall only consider the
regime of parameters for which the corresponding ANO
vortices (without currents) under consideration are of type II
\cite{Vachaspati:2015cma} and restrict attention to the case $n=1$ for
simplicity.

\section{Small condensates}

To motivate the existence of excited solutions, let us first consider
the regime where the condensate is sufficiently small that we can
neglect the term $\propto f^3$ in \eqref{condensate} as well as the
back-reaction on $h$, i.e. the term $\propto f^2$ in \eqref{higgs}.
Assuming the currentless ANO vortex to be stable (hence
choosing $n=1$), we follow the analysis of Witten \cite{witten} and
perturb the scalar field $\sigma$ around $\sigma(x) \equiv 0$ in the
background of this vortex. This will tell us whether in a given setting
the vortex can sustain the additional structure.

Taking the boundary conditions into account, we assume the following
simple ansatz for the Higgs field function\footnote{We have repeated
the calculations with $h(x) = \tanh(\kappa x)$ with qualitatively
similar results \cite{hmp1}.}
\begin{equation}
h(x) = \left\lbrace \begin{array}{lll} \kappa \, x & {\rm for} & 0 \leq
x < 1 / \kappa, \\ 1 & {\rm for} & 1 / \kappa \leq x, \end{array}
\right.
\end{equation}
where $\kappa$ is a freely adjustable constant giving the characteristic
size of the Higgs field variations. For $x \geq 1/\kappa$, equation
\eqref{condensate} then becomes 
\begin{equation}
\frac{1}{x} \lp x \, f' \rp' = \left( w - \gamma_2 \, q^2 +
\gamma_3 \right) f.
\end{equation}
This is a modified Bessel equation whose only solution decaying strictly
faster that $x^{-1/2}$ at infinity is
\begin{equation}
f(x) = C_1 K_0 \left( \sqrt{w-\gamma_2 q^2 + \gamma_3} \, x \right)  \ ,
\end{equation}
where $C_1 \in \mathbb{R}$ is an integration constant and $K_0$ is the
zeroth order modified Bessel function of the second kind. In the
interior region $x < 1/\kappa$, equation \eqref{condensate} becomes
\begin{equation}
\label{eq:lin_int}
\frac{1}{x} \lp x \, f' \rp' = \left( w - \gamma_2 \, q^2 +
\gamma_3 \, \kappa \, x \right) f.
\end{equation}
To find its solutions, it is useful to define the variable $Y(x)$ and the
function $F$ by
\begin{equation}
 Y(x)\equiv \sqrt{\gamma_3} \kappa x^2, \  \ F\left[Y(x)\right] =
 \exp\left[\frac{Y(x)}{2}\right]f(x).
\end{equation}
Equation~\eqref{eq:lin_int} is then rewritten as the linear, second
order differential equation
\begin{equation} 
Y \frac{\dd^2 F}{\dd Y^2} + (1-Y) \frac{\dd F}{\dd Y} + m \, F = 0  \ ,
\label{hyper}
\end{equation}
where 
\begin{equation}
m  \equiv - \left( \displaystyle\frac{w-\gamma_2 q^2}{4
\sqrt{\gamma_3} \kappa} + \frac{1}{2} \right) \ .
\label{wm}
\end{equation}
Eq.~\eqref{hyper} is the confluent hypergeometric
equation~\cite{Abramowitz}. It has only one regular solution at the
origin, namely $F(Y)= C_2 L_{m} (Y)$, where $L_{m}$ denotes the
Laguerre function with parameter $m $ and $C_2 \in \mathbb{R}$ is
another integration constant. So, for $x < 1/\kappa$, the only regular
solution of~\eqref{eq:lin_int} is
\begin{equation}
f(x) = C_2 \, \ex^{- \sqrt{\gamma_3} \kappa x^2 / 2} L_{m
}(\sqrt{\gamma_3} \kappa x^2), \, C_2 \in \mathbb{R}.
\label{eq:fLaguerre}
\end{equation}
The requirement that $f$ and $f'$ should be continuous at $x =
1/\kappa$ gives two matching conditions. The first one is a relation
between $C_1$ and $C_2$. The other one gives  a constraint on $m $ with
a discrete set of solutions (see \cite{hmp1} for more details.) That
is, regular solutions decaying sufficiently fast at infinity exist only
for these discrete values of $m$, which was to be expected
since the linearized version of Eq.~\eqref{condensate} solved here is
nothing but a two dimensional Schr\"odinger equation in a confining
potential with a finite number of bound states.

\begin{figure}
\begin{center}
\includegraphics[width=0.95\linewidth]{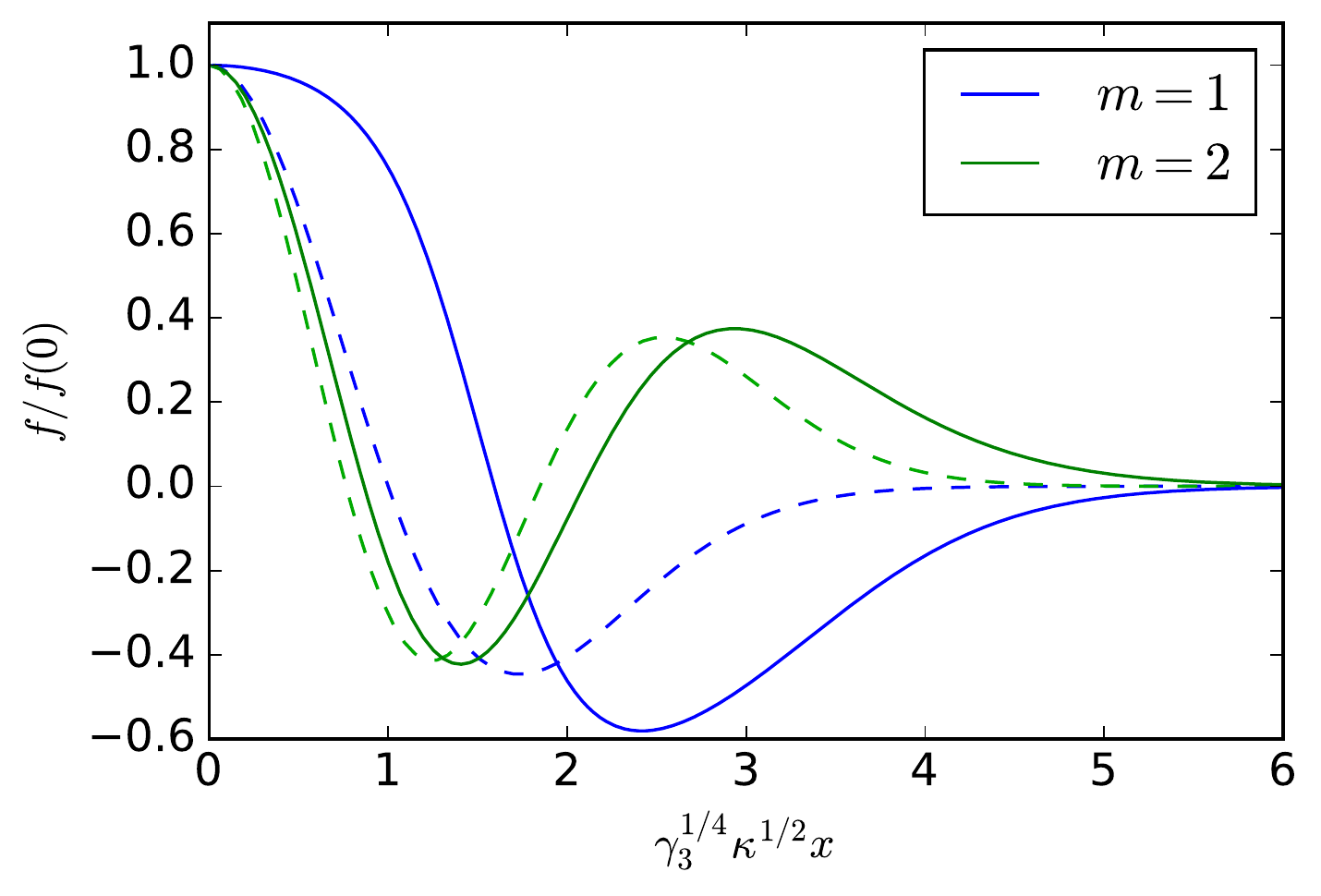}
\end{center}
\caption{Solutions with $m = 1$ and $2$ in the chiral limit $w=0$. The
parameters are $\gamma_2=5000$, $\gamma_3=100$, $q=0.1$ and
$\alpha=10^{-2}$. Dashed lines show the corresponding solutions in the
small condensate approximation for $\sqrt{\gamma_3} / (4 \kappa) \gg 1$,
see Eq.~\eqref{eq:fLaguerre}. Under this assumption, $f / f(0)$ is, for
each value of $m$, a universal function of the rescaled coordinate
$\gamma_3^{1/4} \kappa^{1/2} x$ here used. Although the qualitative
behavior of the solution is well captured by the perturbative
approximation, the actual behavior requires the full numerical solution.
One can also check that the behavior close to the string core ($r\to 0$)
corresponds to the expected $1-f/f(0)\propto r^2$
}
\label{fig:Laguerre1}
\end{figure}

In the limit $\sqrt{\gamma_3} / \kappa \to \infty$, these solutions are
localized in the domain $x \ll 1 / \kappa$, so that \eqref{eq:fLaguerre} is
valid up to a region where $f$ is exponentially small. Regular solutions are
then given in terms of the Laguerre polynomials~\cite{Abramowitz},
restricting $m$ to be a natural integer. As the $m^{\rm th}$ Laguerre
polynomial has $m$ strictly positive roots, the corresponding function $f$
has $m$ nodes (see Fig.~\ref{fig:Laguerre1}, dashed lines for $m = 1$ and $m =
2$). In practice, the finite value of $\kappa$ gives a cutoff on $m$, of the
order of $\sqrt{\gamma_3} / (4 \kappa)$. For the numerical solutions shown in
the figure, we estimate that $\sqrt{\gamma_3} / (4 \kappa) \approx 4.90$ for
$m=1$ and $\sqrt{\gamma_3} / (4 \kappa) \approx 4.32$ for $m=2$,
respectively\footnote{The difference between these two values is due to the
back-reaction of the condensate on the Higgs field, which is not taken into
account in the linear analysis. While this is a limitation on the relevance of
the bound on $m$, the latter is still expected to give the correct order of
magnitude.}. Since nonlinear effects tend to reduce the number of solutions, we
thus expect to have a few of them when working with the full set of equations.
This is compatible with the fact that for these parameters, only solutions
with $0$, $1$, and $2$ nodes exist, see the next section.
In the following, for simplicity we denote by $m$ the number of nodes of the
function $f$, although it does not exactly follow \eqref{eq:fLaguerre}.

The existence of bound states around an otherwise stable ANO vortex is the
reason why an ordinary cosmic string can become superconducting \cite{witten}.
Our discussion here goes one step further by showing that there is a finite
number of such bound states, each of which leading to an instability in the
current condensate, an instability that will be tamed by the
nonlinear terms: one therefore expects that for each value of $w$, the full
nonlinear set of equations should lead to a series of solutions with finite
energy per unit length and tension in the form of a ground state and excited
modes.

Indeed, so far, we have worked to linear order in $f$. When including
the nonlinear term $\propto f^3$ and its backreaction on $h$ and $P$,
the amplitude of the regular, asymptotically decaying solutions depends
on the values of $w$. Since the boundary conditions and asymptotic
behaviors of bounded solutions remain similar to the linear case, one
can conjecture they remain discrete, and can be continuously deformed
into the linear solutions, so it seems reasonable to assume that the
qualitative features of the solutions do not change. This results in a
discrete set of series of solutions, each of them extending over a
finite interval of $w$. We present these solutions numerically in the
next section. In Fig.~\ref{fig:f0w}, we present the dependence of $w$
on $f(0)$ due to the nonlinear and backreaction terms.

\section{Numerical results}
\begin{figure}
\begin{center}
\includegraphics[width = 0.95\linewidth]{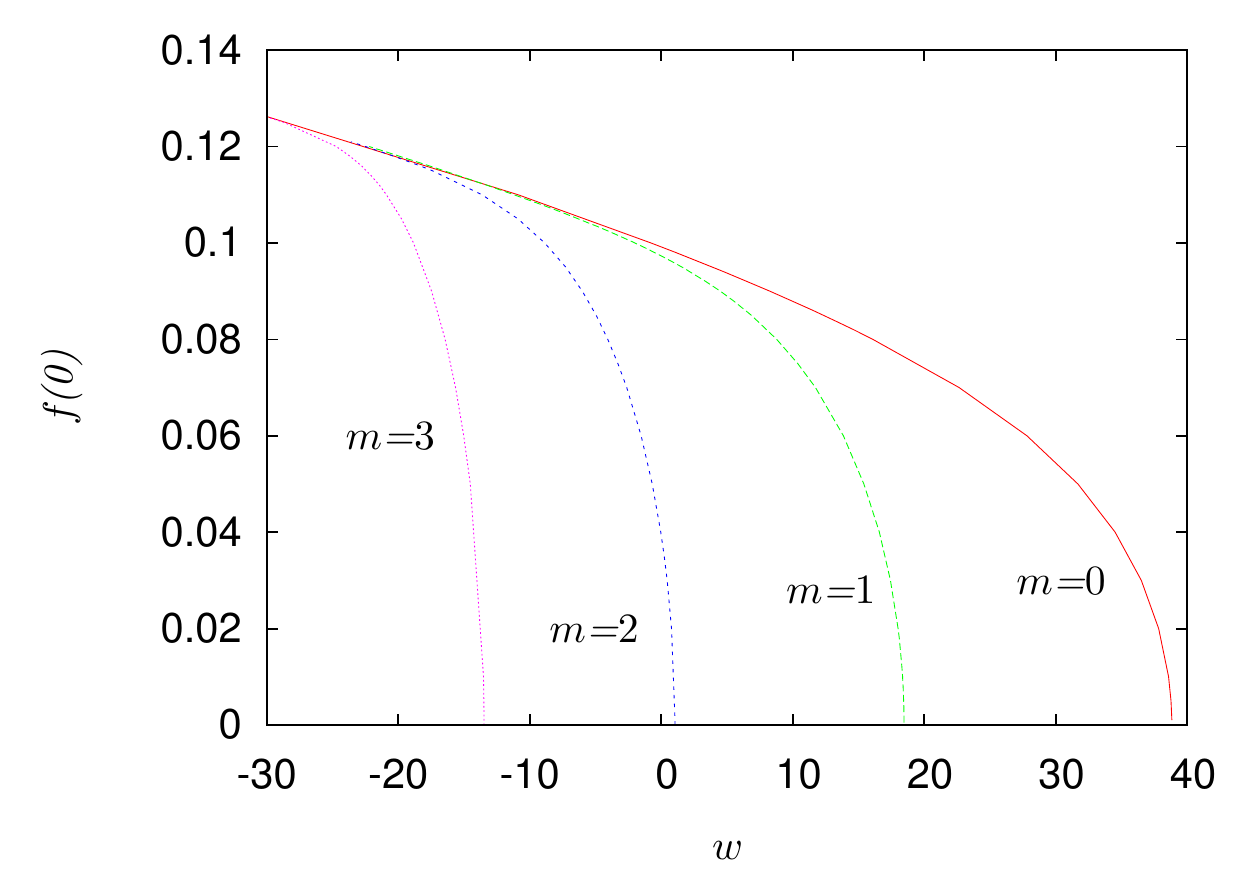} 
\end{center}
\caption{Value $f(0)$ of the condensate function along the string axis
as a function of the state parameter $w$ for excited state numbers
$m=0,1,2,3$ and the same parameters as in Fig.~\ref{fig:Laguerre1}.}
\label{fig:f0w}
\end{figure}

We have solved numerically the set of coupled non-linear differential
equations (\ref{gauge}), (\ref{higgs}), and (\ref{condensate}), subject
to the appropriate boundary conditions \eqref{BoundCond} using an
iterative Newton-Raphson/collocation method with automatic grid
selection \cite{colsys}. The relative errors of our solutions are
typically of the order of $10^{-7}$ to $10^{-10}$. We have studied a
large number of parameter values, $\alpha \in [10^{-3}, 10^{-1}]$,
$\gamma_i \in [1, 10^3]$, $i=2,3$ and $q \in ]0,1]$, but report only on
one specific case that is generic enough and displays the main
qualitative features. The choice was made in order to fulfill the
necessary constraints (see e.g. \cite{hartmann_carter}) by a large
margin and to allow the existence of solutions in the chiral limit
$w=0$. The chiral solutions with $m=1,2$ nodes for our choice of
parameters are shown in Fig.~\ref{fig:Laguerre1}, together with the
approximate perturbative solutions. The constant $\kappa :=
h'\vert_{x=0}$ is extracted from the numerical data. Our numerical
results show that for this choice of parameters chiral solutions can
only possess up to 2 nodes, i.e., $m \leq 2$. This is shown in
Fig.~\ref{fig:f0w}, where we give the value of $f(0)$ as function of the
state parameter $w$. This shows that, while for node number $m=0,1,2$
electric, magnetic as well as chiral solutions exist, $m=3$ solutions
are always electric.

\begin{figure}[h!]
\begin{center}
\includegraphics[width = 0.95\linewidth]{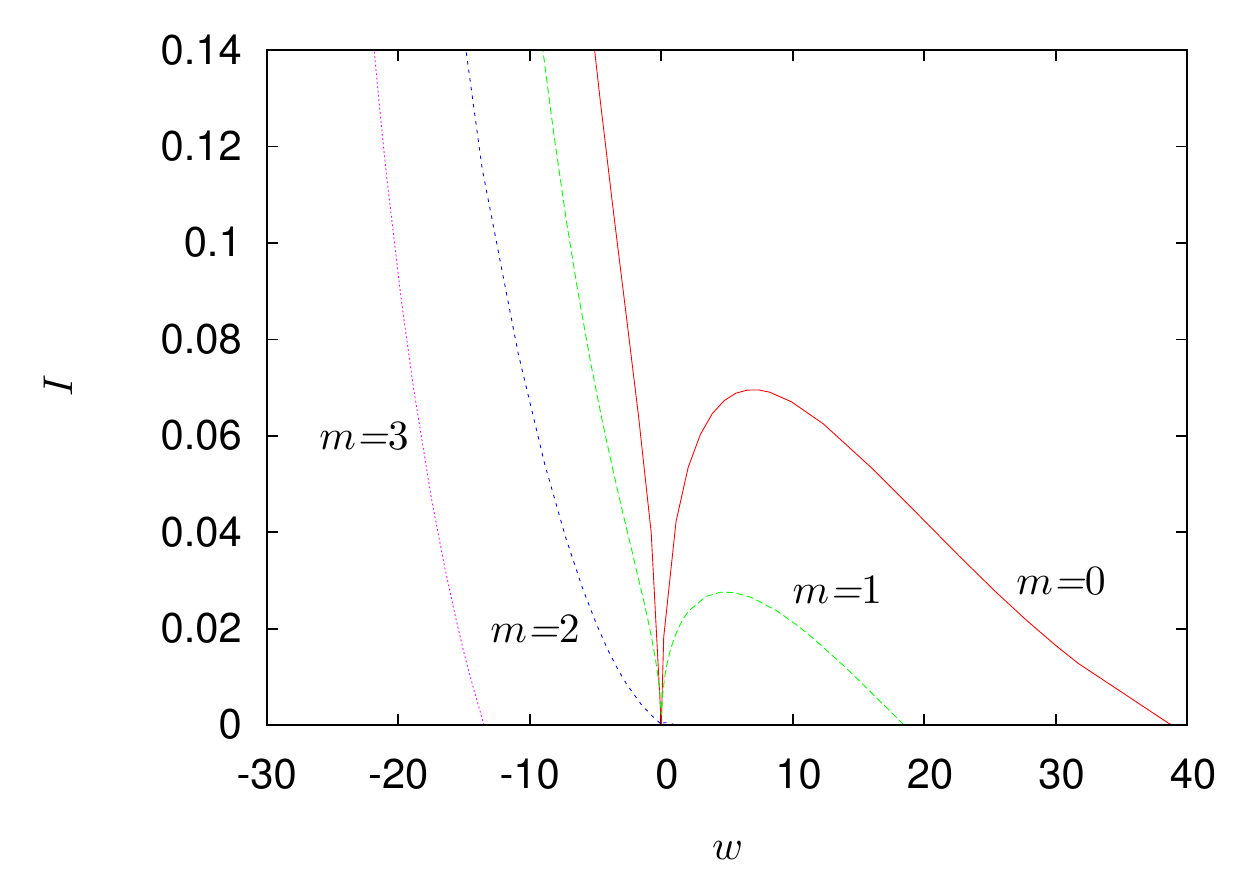} 
\end{center}
\caption{Current $I$ as a function of the state parameter $w$ and the
same parameters as in Fig.~\ref{fig:Laguerre1}.}
\label{fig:current}
\end{figure}

\begin{figure}[h!]
\begin{center}
\includegraphics[width = 0.95\linewidth]{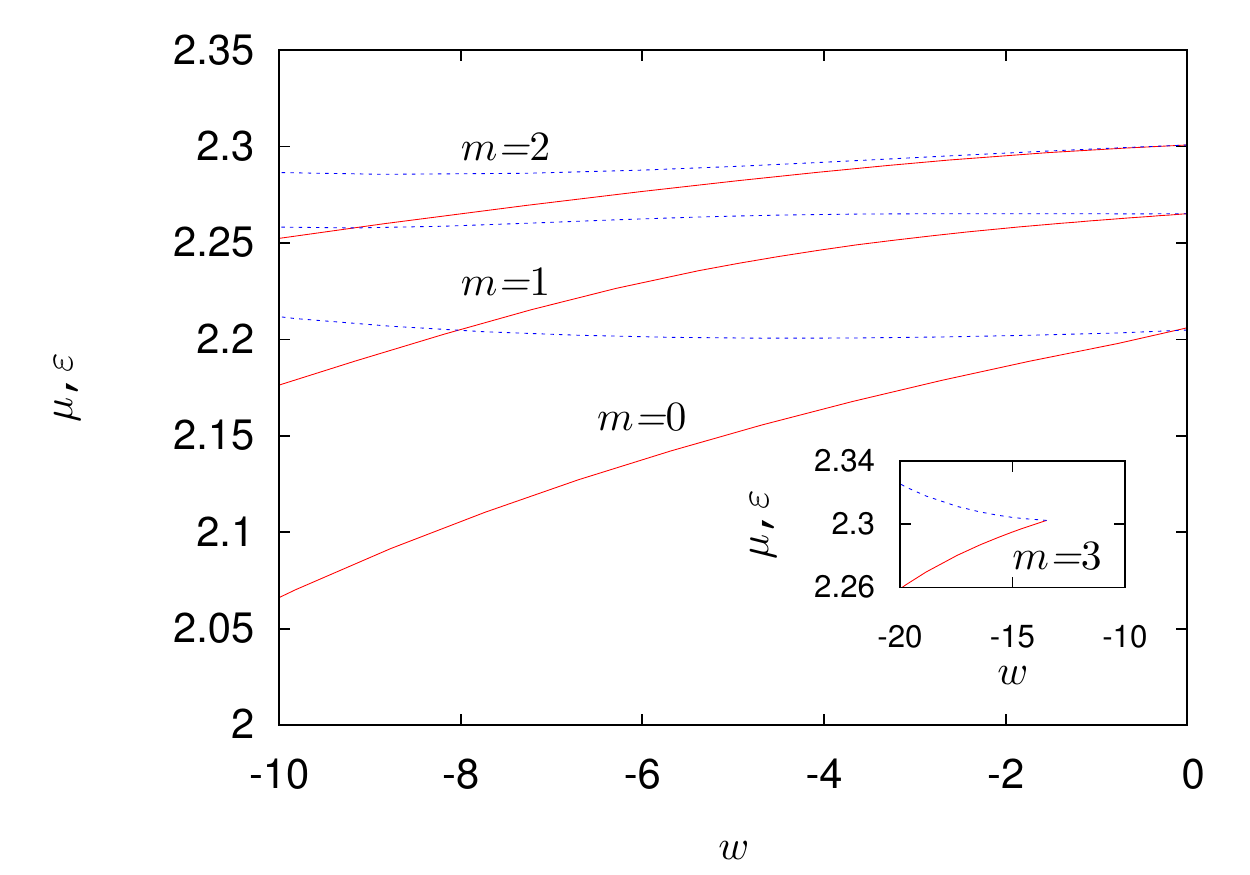} 
\includegraphics[width = 0.95\linewidth]{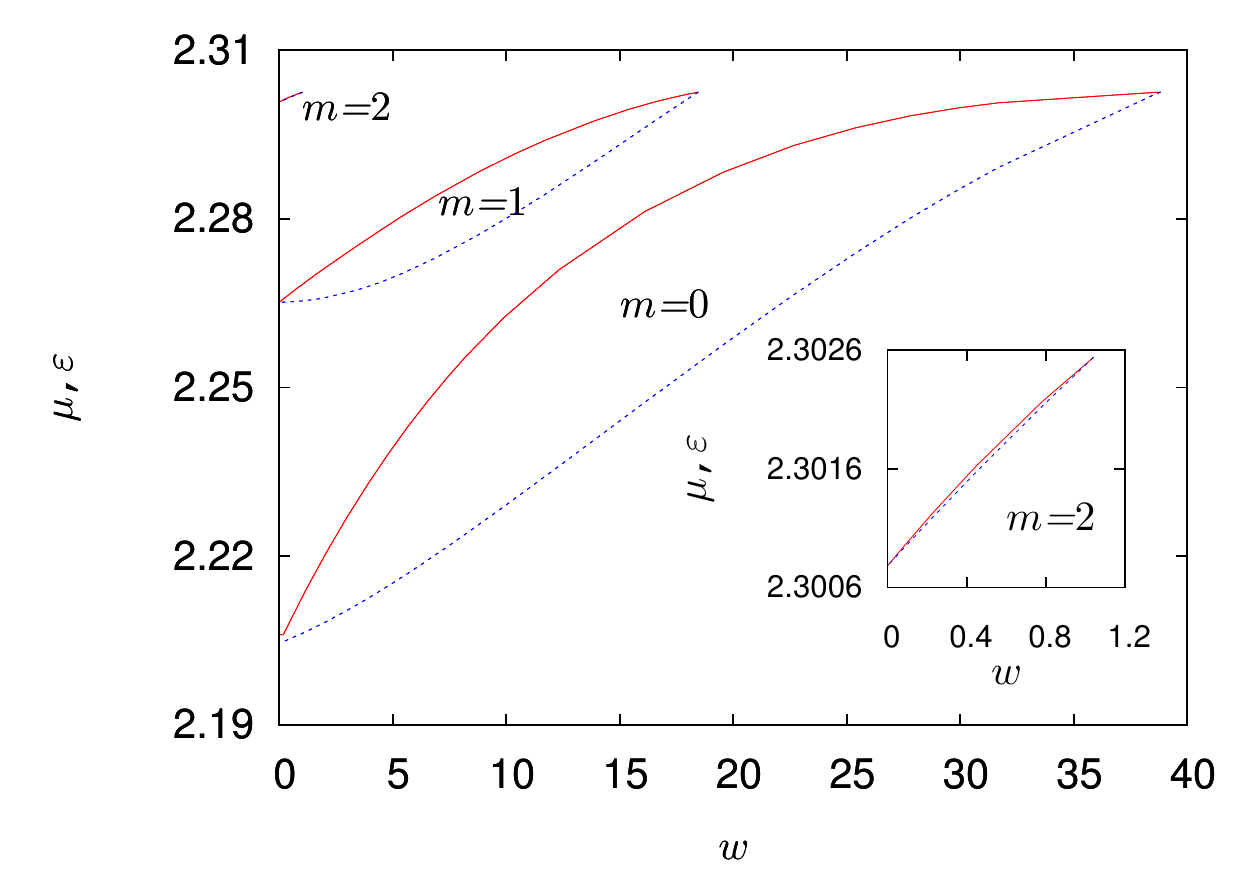} 
\end{center}
\caption{Energy per unit length $\varepsilon$ (red solid) and tension
$\mu$ (blue dashed) as functions of the state parameter $w$ and for the
same parameters as in Fig.~\ref{fig:Laguerre1} in the electric regime
(top) and in the magnetic regime (bottom), respectively.}
\label{fig:mu_eps}
\end{figure}

Fig.~\ref{fig:current} displays the current $I$ of the superconducting
string. We observe that the maximal possible current on the string in
the magnetic regime decreases with increasing node number. In
Fig.~\ref{fig:mu_eps}, we show the energy per unit length $\varepsilon$
as well as the tension $\mu$ of the strings. The qualitative behaviour
of these quantities is similar in all cases.

Carter devised a macroscopic stability criterion \cite{carter} stating
that a necessary condition for stability of superconducting strings is
that the velocities of longitudinal ({\small{L}}), $c_{_\mathrm{L}}$,
and transverse ({\small{T}}) perturbations, $c_{_\mathrm{T}}$, given by
\begin{equation} 
c_{_\mathrm{L}} =\sqrt{-\frac{\dd\mu}{\dd \varepsilon}}, \ \
c_{_\mathrm{T}}=\sqrt{\frac{\mu}{\varepsilon}},
\end{equation}
respectively, should be real for the string to be stable. This occurs
for the state parameter less than a limiting value (in general different
for each series of solutions) which we denote by $w_\mathrm{Carter}$. As
can be seen from Fig.~\ref{fig:mu_eps}, in our example we have
$w_{\textrm{Carter}} \approx -4$ for $m=0$, and $w_{\textrm{Carter}}
\approx-9$ for $m = 1,2$. We expect that $w$ can become positive for a
different choice of parameters, as demonstrated for $m = 0$
in~\cite{Peter:1992dw}.

This criterion being a necessary condition, it permits to restrict the
range of parameters in which stable solutions may exist, but doesn't
guarantee stability, even to linear order: they may support unstable
modes not captured by this macroscopic criterion. Determining the
presence or absence of such modes requires linearizing the field
equations from the Lagrangian~\eqref{lag}. A full linear stability
analysis is beyond the scope of the present letter and will be
presented in~\cite{hmp1}. Here we exemplify the point by focusing on
perturbations of $f$ only, keeping the other fields fixed.~\footnote{A
comment about self-consistency is in order here. Strictly speaking, the
stability analysis should be done by considering variations of the
three fields independently, as they are all related by the field
equations, and it is not necessarily clear \textit{a priori} what can
be learned by considering perturbations of one field only. However, one
can show that in the present case the existence of an instability at
fixed $h$ and $P$ implies that the full system is unstable (although
the converse may not be true). The proof will be given in~\cite{hmp1},
along with a more detailed and complete stability analysis.}

\begin{figure}[h!]
\begin{center}
\includegraphics[width = 0.9\linewidth]{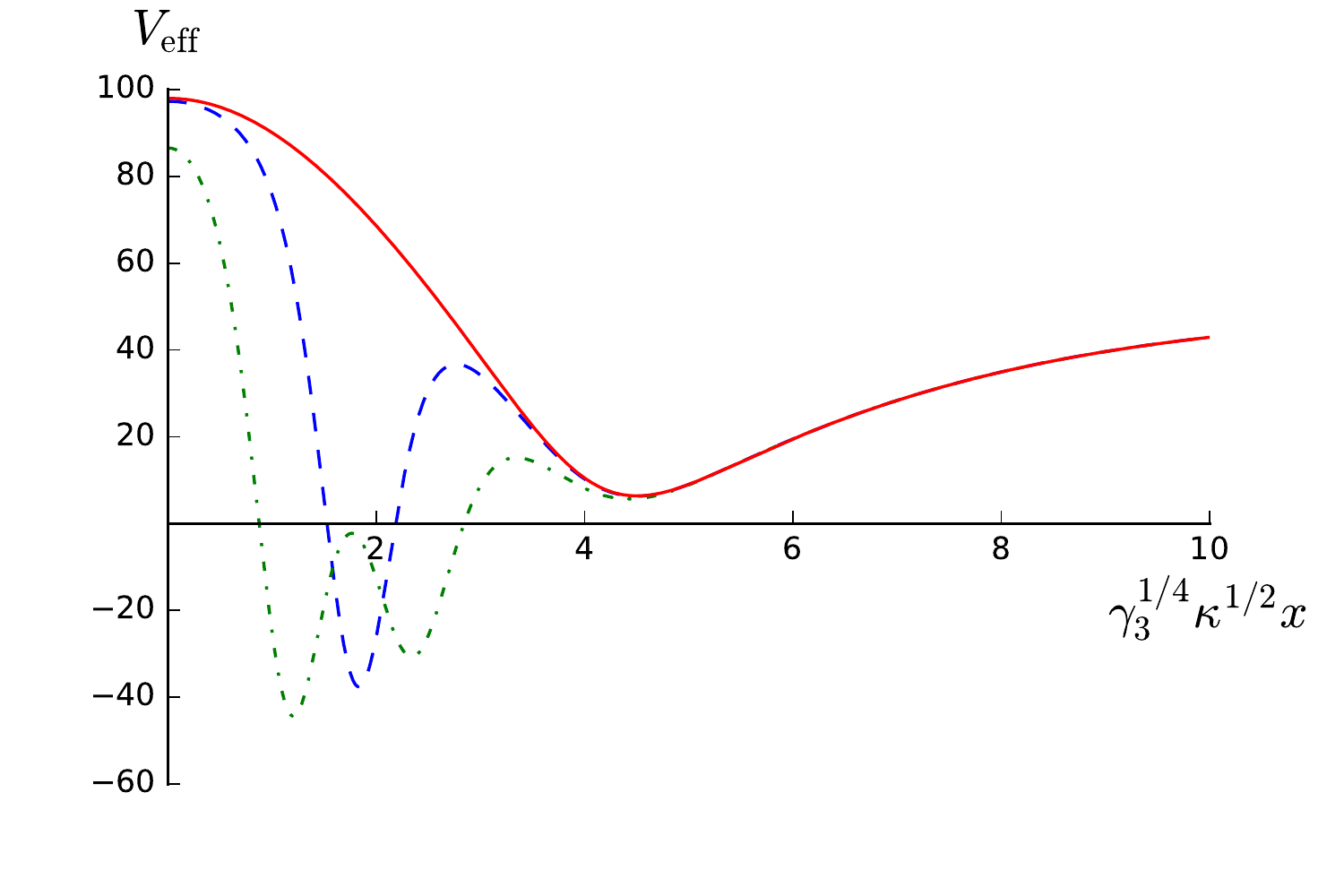}
\includegraphics[width = 0.9\linewidth]{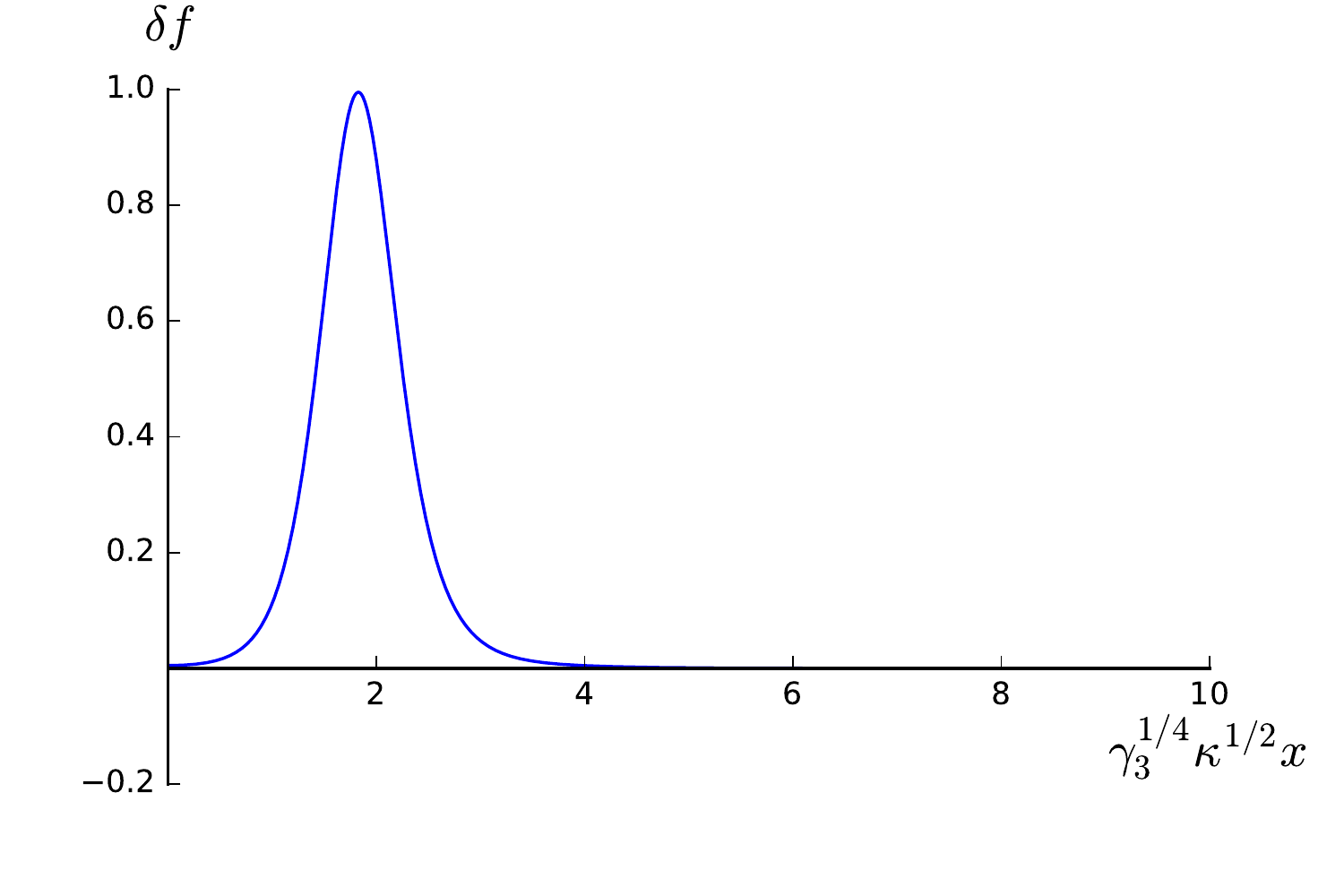}
\includegraphics[width = 0.9\linewidth]{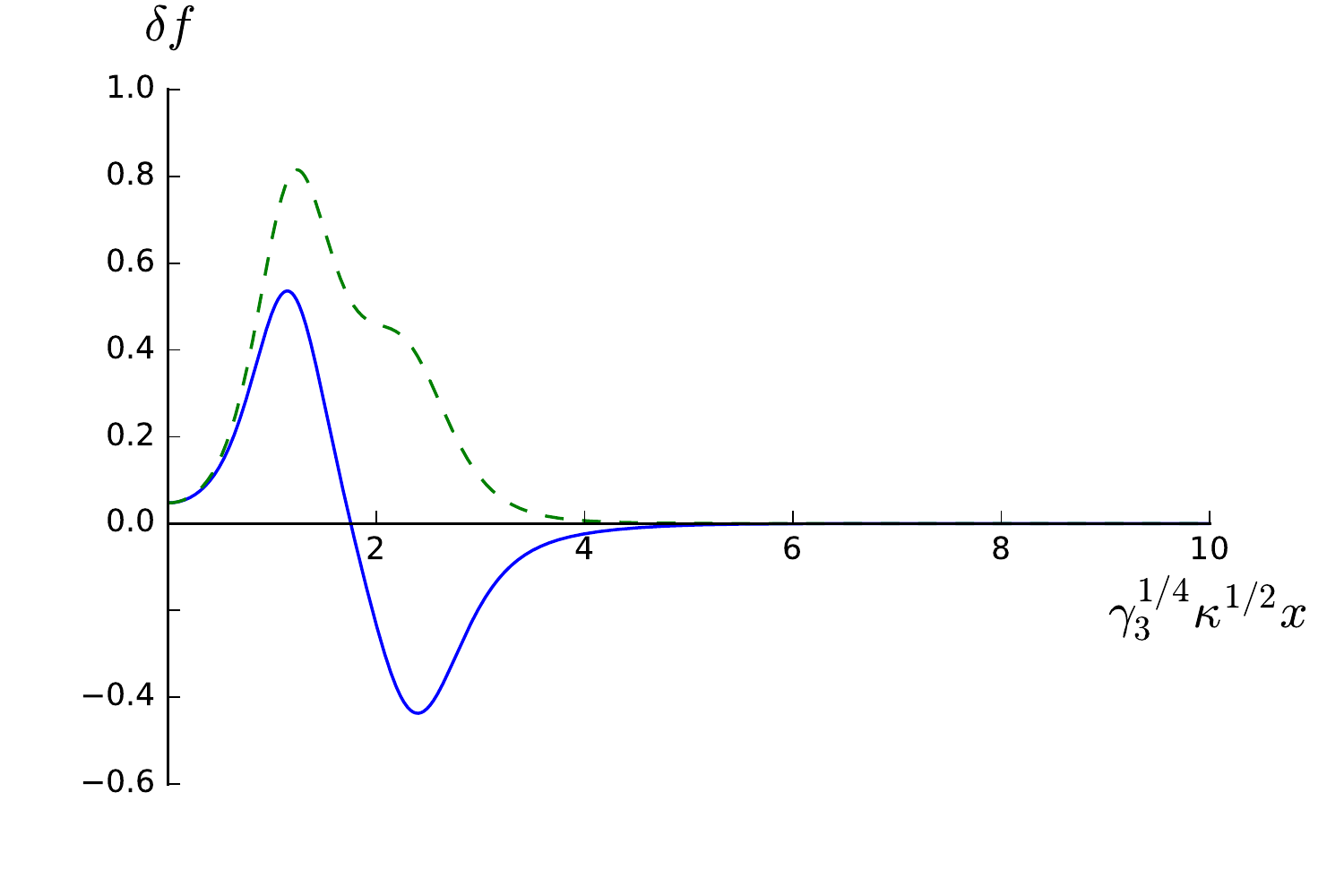}
\end{center}
\caption{Top panel: Effective potential $V_\textrm{eff}$ for the
solutions with $m = 0$ (red, continuous), $m=1$ (blue, dashed), and
$m=2$ (green, dot-dashed) in the small-condensate limit. The profile of
$h$ is a hyperbolic tangent with parameters chosen to match those of
the numerical solution with $m = 0$. Middle panel: profile of the
unstable mode for the solution with $m=1$. (The normalization is
arbitrary.) Bottom panel: profiles of the two unstable modes for the
solution with $m=2$.}
\label{fig:Veff}
\end{figure}

Inserting $f(x)=f_0+\delta f$, where $f_0(x)$ is a solution to the full
set of equations (\ref{gauge}), (\ref{higgs}), (\ref{condensate}) and
$\delta f(x)$ is real and small, we find that (\ref{condensate})
becomes -- neglecting terms of order $(\delta f)^2$ and higher --  a
Schr\"odinger-type equation for $\delta f$ with potential
$V_\textrm{eff}(x)=w+3\gamma_2f_0^2 -\gamma_2 q^2+\gamma_3 h_0^2$,
where $h_0(x) $ is a solution to (\ref{gauge}), (\ref{higgs}),
(\ref{condensate}). We find that for the solutions plotted in
Fig.~\ref{fig:Laguerre1}, this potential is strictly positive for the
fundamental solution, but becomes negative in some range of the
coordinate $x$ for the excited solutions for $w \leq 0$. This indicates
a possible instability. Solving numerically the time-dependent equation
on $\delta f$, we found that an unstable mode indeed exists. This is
illustrated in Fig.~\ref{fig:Veff}, showing $V_\textrm{eff}$ for the
fundamental and the two excited solutions computed in the
small-condensate limit as well as the corresponding unstable modes.
Since an instability already shows up in this linearized model, one
should envisage that most of the excited solutions may be unstable.
The question remains to what extent this is a generic
(parameter-dependent) statement. Moreover, in the magnetic regime for
which $w>0$, there must exist a threshold $w_\mathrm{th}$ above which
the potential becomes positive definite again. The relevant solutions
may then be stable provided $w_\mathrm{th}$ is below
$w_\mathrm{Carter}$ defined above.

\section{Discussion}
\label{sec:discussion}

The above strings are of the purely spatially local type as far as the
current is concerned, and this may, in some GUT-based models for
instance, be considered unlikely, as most symmetries then are expected
to be gauged. Indeed, many superconducting cosmic string models
consider actual electromagnetic currents, and may even be used to
produce large-scale primordial magnetic fields. The excited states
found here are based on an uncoupled model, but we have constructed the
corresponding solutions in the case where $\sigma$ is given a gauge
coupling too, i.e. for $e_\sigma\not= 0$, and found that the basic
features are left unchanged \cite{hmp1}. Besides, as discussed in
Ref.~\cite{enon0}, the influence of such an extra coupling is expected
to be mostly negligible on the microstructure and we confirmed this
expectation for the excited solutions. On the other hand, such a
coupling could source a different external field, leading to a more
complicated large scale magnetic structure; this should be considered
in a cosmological context.

We would also like to remark that the main qualitative features of
these solutions persist in a dynamical space-time, i.e. when coupling
the matter equations to the Einstein equation. We have checked this
numerically and will report on the details of our results elsewhere
\cite{hmp1}.

We find some of these solutions to fulfill the macroscopic stability
criterion of Carter, which uses integrated, macroscopic quantities
regardless of the details of the underlying field theoretical model.
In contrast, the {\it microscopic stability} analysis of these solutions
is very much model-dependent.
Since the integrated quantities stem from the underlying model though, we expect that
a solution that is {\it macroscopically unstable} will also be unstable when investigating the micro-physics. Our preliminary results indicate that
the excited solutions may be classically unstable {\it microscopically}, so we should not expect
them to be directly observable through, e.g., gravitational lensing. Besides,
it would be very difficult to distinguish an excited from a standard
cosmic string through the deficit angle alone. On the other hand, the fact that
these solutions are unstable may offer a possibility to detect them
through particle emission during the collision and recombination of two
strings. For instance, the large fluctuations of the fields during the
collision may locally excite the condensate, producing a solution with
$m \neq 0$ along a segment of one or both strings, the sharp transition
regions between the parts with $m=0$ and $m \neq 0$ producing a current
kink propagating along the string axis. Since the solutions with $m \neq
0$ are unstable, they will locally decay to the fundamental one,
emitting high-energy particles. One can thus expect radiation with high
energy to be emitted along a one-dimensional region of space, following
the trajectory of the kink. In return, this emission should reduce the
average velocity of the string network, transforming its kinetic energy
into massive particle radiation.

Interestingly, the excited solutions may have close analogues in
Bose-Einstein condensates, which would open possible paths to
observations in laboratory experiments. It is now well
known~\cite{Vortices} that rotating condensates develop vortex lines
akin to cosmic strings. Let us consider 
a cold gas made of two
different atoms, or a single species of atoms with two internal states
coupling differently to some external electromagnetic fields creating
the confining potential. %One can imagine that 
In principle,
the potentials
experienced by the two types of atoms %, their interactions with each
%others, and the temperature 
can be tuned so that only one type of atoms
condenses in the lowest-energy state, while the other type can condense
inside a vortex core. %where the density of the first type is small. 
The vortex could then support a supercurrent, exactly like the strings
considered in this letter. Moreover, from the similarity between
\eqref{gauge} and the stationary Gross-Pitaevskii equation, one can
expect to find the same structure of solutions, with one fundamental
configuration minimizing the energy and, depending on the parameters, a
set of excited ones with a more complicate structure for the condensate
of the second type of atoms. (%As mentioned above, 
This will be true generically provided the potential experienced by the second
type of atoms is quadratic close to the vortex line.) To the best of our
knowledge, the question of whether such a setup is possible or not is still pending.

\section*{Acknowledgement} 

BH would like to thank Brandon Carter for discussions during the initial
stages of this work. BH would like to thank CNPq for financial support
under {\it Bolsa de produtividade Grant No. 304100/2015-3}. BH and PP
would like to thank FAPESP for financial support under {\it Project No.
2015/02563-8}. PP would like to thank the Labex Institut Lagrange de
Paris (reference ANR-10-LABX-63) part of the Idex SUPER, within which
this work has been partly done.
\\ \\

\end{document}